\documentclass[aip,jcp,reprint,noshowkeys,superscriptaddress]{revtex4-1}
\usepackage{graphicx,dcolumn,bm,xcolor,microtype,multirow,amscd,amsmath,amssymb,amsfonts,physics,wrapfig,txfonts,siunitx}
\usepackage[version=4]{mhchem}
\newcommand{\alert}[1]{\textcolor{black}{#1}}
\usepackage[normalem]{ulem}

\newcommand{\SupInf}{\textcolor{blue}{Supporting Information}}

\newcommand{\mc}{\multicolumn}
\newcommand{\fnm}{\footnotemark}
\newcommand{\fnt}{\footnotetext}
\newcommand{\QP}{\textsc{quantum package}}

\usepackage[
	colorlinks=true,
    citecolor=blue,
    breaklinks=true
	]{hyperref}
\begin{document}

\newcommand{\LCPQ}{Laboratoire de Chimie et Physique Quantiques (UMR 5626), Universit\'e de Toulouse, CNRS, UPS, France}

\title{State-Specific Configuration Interaction for Excited States}

\author{F\'abris Kossoski}
\email{fkossoski@irsamc.ups-tlse.fr}
\affiliation{\LCPQ}
\author{Pierre-Fran\c{c}ois Loos}
\email{loos@irsamc.ups-tlse.fr}
\affiliation{\LCPQ}

\begin{abstract}
{\bf Abstract:}
We introduce and benchmark a systematically improvable route for excited-state calculations, state-specific configuration interaction ($\Delta$CI),
which is a particular realization of multiconfigurational self-consistent field and multireference configuration interaction.
Starting with a reference built from optimized configuration state functions, separate CI calculations are performed for each targeted state (hence state-specific orbitals and determinants).
Accounting for single and double excitations produces the $\Delta$CISD model, which can be improved with second-order Epstein-Nesbet perturbation theory ($\Delta$CISD+EN2) or a posteriori Davidson corrections ($\Delta$CISD+Q).
These models were gauged against a vast and diverse set of 294 reference excitation energies.
We have found that $\Delta$CI is significantly more accurate than standard ground-state-based CI, whereas close performances were found between $\Delta$CISD and EOM-CC2, and between $\Delta$CISD+EN2 and EOM-CCSD.
For larger systems, $\Delta$CISD+Q delivers more accurate results than EOM-CC2 and EOM-CCSD.
The $\Delta$CI route can handle challenging multireference problems, singly- and doubly-excited states, from closed- and open-shell species, with overall comparable accuracy,
and thus represents a promising alternative to more established methodologies.
In its current form, however, it is only reliable for relatively low-lying excited states.
\end{abstract}

\maketitle

\section{Introduction}
\label{sec:intro}

Most molecular electronic structure methods rely on different descriptions for ground and excited states.
The ground state is described first, at a given level of theory, providing a baseline for later accessing the excited states, which in turn makes use of another approach or a distinct formalism altogether.
For example, Kohn-Sham (KS) density-functional theory (DFT) is a ground-state method, \cite{Hohenberg_1964,Kohn_1965,Parr_book}
whereas the excited states are obtained later with a linear response treatment of time-dependent density-functional theory (TDDFT). \cite{Runge_1984,Burke_2005,Casida_2012,Huix-Rotllant_2020}
Similarly, the coupled-cluster (CC) \cite{Cizek_1966,Crawford_2000,Bartlett_2007,Shavitt_2009} equations are solved for the ground state,
whereas a diagonalization of the similarity-transformed Hamiltonian is implied in excited-state calculations based on the equation-of-motion (EOM) \cite{Rowe_1968,Stanton_1993} or linear-response \cite{Monkhorst_1977,Koch_1994} formalisms.
Within configuration interaction (CI) methods, \cite{Szabo_book} the underlying formalism is the same for ground and excited states, but typical implementations also rely on a special treatment for the ground state,
given the use of ground-state Hartree-Fock (HF) orbitals and the fact that the truncated CI space is spanned by excitations from the ground-state HF determinant.

Whereas the above-mentioned methods rely on a single determinant reference, enlarging the reference space with more than one determinant gives rise to multi-reference approaches.
In multiconfigurational self-consistent field (MCSCF), \cite{Das_1966,Roos_1980,Roos_1980a,book_multiconfigurational}
the wave function is expanded as a linear combination of an arbitrary set of determinants,
and the orbitals (and the coefficients of these determinants) are optimized to make the energy stationary.
The most employed type of MCSCF is complete active space self-consistent field (CASSCF), \cite{Roos_1980}
which allows for all determinants generated by distributing a given number of electrons in a given number of active orbitals.
Multireference CI (MRCI) offers a route to go beyond MCSCF by considering excited determinants generated from the reference space,
which in practice is limited to single and double excitations (MRCISD).
The MRCISD energy can be further improved with so-called Davidson corrections. \cite{Langhoff_1974,Pople_1977,Szalay_2012}

Apart from multireference approaches, \cite{Szalay_2012,Lischka_2018} single-reference excited state methods entail a formal distinction between the targeted excited states and the ground state.
It is thus important to devise methods that minimize this unbalance as much as possible, aiming at a more unified description of ground and excited states, while keeping a modest computational cost.
This also means a more balanced description among the excited states, and here we highlight the case of singly- and doubly-excited states, which differ by the number of excited electrons during the electronic transition.
Most excited-state methodologies either fail to properly describe doubly-excited states or require higher-order excitations to be accounted for. \cite{Loos_2019}
In this sense, a methodology that offers a comparable accuracy for singly- and doubly-excited states would be equally desirable.

MCSCF methods can be either 
state-averaged, when the reference space is optimized for an ensemble of (typically equally weighted) states,
or state-specific, when the optimization is performed for one targeted state.
The state-averaged strategy is much more used in practice,
mostly because of the more straightforward and reliable orbital optimization and the easier calculation of transition properties (given the common set of orbitals),
when compared to the state-specific approach.
\cite{Das_1973,Dalgaard_1978,Lengsfield_1980,Bauschlicher_1980,Bauschlicher_1980a,Werner_1981,Golab_1983,Olsen_1983} 
However, state-averaged MCSCF faces several issues.
It struggles to describe higher-lying states or a large number of states, the orbitals may favor some states in the detriment of others, \cite{Meyer_2014,Segado_2016,Tran_2019,Tran_2020}
the potential energy curves can become discontinuous, \cite{Zaitsevskii_1994,Glover_2014,Tran_2019} and the calculation of energy derivatives is complicated by the energy averaging. \cite{Dudley_2006,Granovsky_2015,Snyder_2017}
Many if not all of these problems do not appear in state-specific MCSCF, 
which in turn has to deal with a more challenging orbital optimization problem.

In light of these motivations, 
there has been an ever-growing interest in state-specific MCSCF, \cite{Shea_2018,Tran_2019,Tran_2020,Burton_2022} and state-specific methods in general.
The general principle is to employ a single formalism, approaching each state of interest independently, and without resorting to any prior knowledge about the other states.
The first and probably the most well-known state-specific method is $\Delta$SCF, \cite{Ziegler_1977,Kowalczyk_2011}
where excited states are described by a single determinant and represent higher-lying solutions of the HF or KS equations.
By optimizing the orbitals for a non-Aufbau determinant, $\Delta$SCF attempts to recover relaxation effects already at a mean-field level.
There is a growing body of evidence showing that DFT-based $\Delta$SCF usually outperforms TDDFT,
\cite{Kowalczyk_2013,Gilbert_2008,Barca_2018,Hait_2020,Hait_2021,Shea_2018,Shea_2020,Hardikar_2020,Levi_2020,Carter-Fenk_2020,Toffoli_2022}
most notably for doubly-excited and charge transfer states. \cite{Hait_2020,Hait_2021}
However, $\Delta$SCF still represents a major limitation to open-shell singlet states, because of strong spin-contamination associated with the single-determinant ansatz.
Restricted open-shell Kohn-Sham (ROKS) \cite{Filatov_1999,Kowalczyk_2013} offers one way around this problem,
by optimizing the orbitals for a Lagrangian that considers both the mixed-spin determinant and the triplet determinant with spin projection $m_s$ = 1.
In wave-function-based methods, excited-state mean field (ESMF) theory \cite{Shea_2018,Shea_2020,Hardikar_2020}
has been proposed as a state-specific MCSCF alternative to excited states.
In the ESMF approach, excited-state orbitals are optimized for a CI with single excitations (CIS) ansatz, \cite{Szabo_book}
and energies can be further corrected with second-order M{\o}ller-Plesset (MP2) perturbation theory. \cite{Shea_2018,Clune_2020}
An extension of ESMF to DFT has also been proposed. \cite{Zhao_2020}
Variants of CC methods that directly target excited states have also been actively pursued. \cite{Piecuch_2000,Mayhall_2010,Lee_2019,Kossoski_2021}

An important practical question for all the aforementioned methods concerns the optimization of orbitals for excited states, which typically appear as saddle point solutions in the orbital parameter space, \cite{Kossoski_2021,Marie_2021,Damour_2021,Burton_2021,Burton_2022}
therefore being more difficult to obtain than ground-state solutions. \cite{Cuzzocrea_2020,Otis_2020,Shepard_2022}
In this sense, specialized algorithms for obtaining excited-state orbitals have been proposed and developed by several groups. \cite{Gilbert_2008,Barca_2018,Hait_2020,Levi_2020,Carter-Fenk_2020}
Related methods that aim at describing multiple states within the same theoretical framework, though usually in a state-averaged fashion, include
CASSCF, \cite{Roos_1980} ensemble DFT, \cite{Cernatic_2022,Gould_2022,Gould_2021,Marut_2020,Loos_2020c,Deur_2019,Gould_2018,Deur_2017,Filatov_2016,Senjean_2015} and multi-state TDDFT. \cite{Gao_2016,Lu_2022,Ren_2016,Horbatenko_2019,Horbatenko_2021}

\section{State-specific CI}
\label{sec:ssCI}

\begin{figure}
\includegraphics[width=1.0\linewidth]{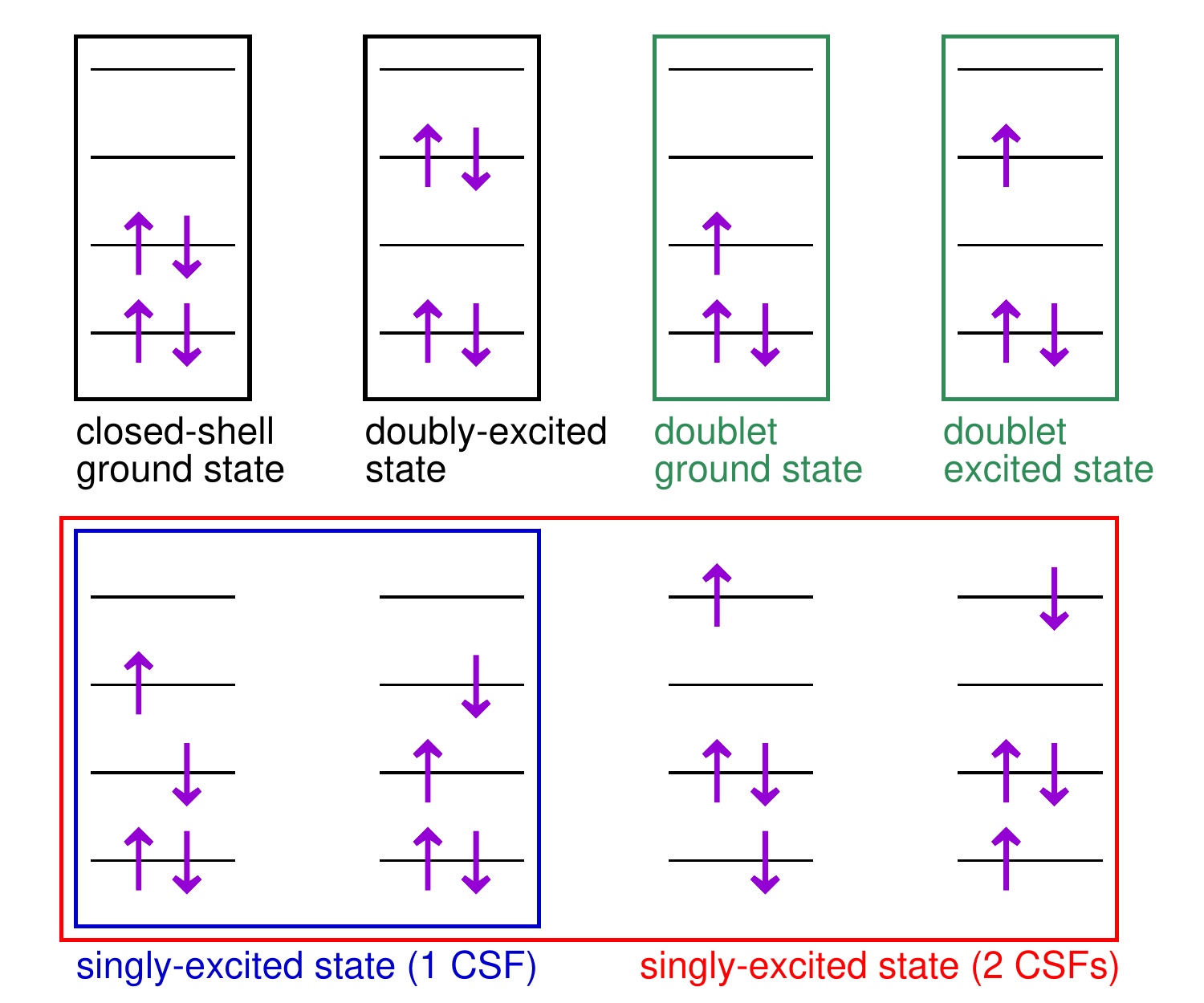}
\caption{Types of configuration state functions (CSFs) employed as a reference for different classes of excited states in our $\Delta$CSF and $\Delta$CISD approaches.}
\label{fig:determinants}
\end{figure}

Here we propose a particular realization of state-specific MCSCF and MRCI as a route for excited-state calculations.
First, the orbitals are optimized at the MCSCF level comprising a minimal set of configuration state functions (CSFs), as illustrated in Fig.~\ref{fig:determinants}, which provides a state-specific reference.
By running separate calculations for the ground state and for a targeted excited state, excitation energies can be obtained as the energy difference between the individual total energies.
We label this approach $\Delta$CSF, in close parallel to the $\Delta$SCF method.
When compared with larger MCSCF choices, the compactness of $\Delta$CSF avoids redundant solutions and is expected to facilitate the convergence toward excited states.
For a single CSF ansatz in particular, the CI coefficients are fixed by the desired spin eigenstate,
eliminating the redundancies associated with the coupling between CI coefficients and orbital rotations.
Furthermore, by being a proper eigenstate of the total spin operator, $\Delta$CSF cures the spin-contamination problem of $\Delta$SCF,
thus leading to truly state-specific orbitals and an improved reference, particularly for singlet excited states.
Finally, being a mean-field method [with an $\order*{N^5}$ computational cost associated with the integral transformation, where $N$ is the number of basis functions],
$\Delta$CSF is intended to provide a balanced set of reference wave functions for a subsequent, more accurate calculation.

At this second stage, correlation effects are captured by performing separate MRCI calculations for each state.
Since ground- and excited-state references are of mean-field quality and are state-specific, this particular type of MRCI calculation is labeled $\Delta$CI here.
When accounting for all single and double excitations, we obtain the $\Delta$CISD model, which is now expected to provide decent excitation energies with an $\order*{N^6}$ computational scaling.
Notice that, because we perform all singles and doubles with respect to each reference determinant, the maximum excitation degree is potentially higher than two (except of course for a one-determinant reference).
This also applies to higher-order CI calculations.
In this way, each state is described as much as possible in a state-specific way, with a different set of orbitals as well as determinants.
\alert{Also notice that, since we aim for a state-specific treatment of correlation,
one cannot anticipate which root of the CI calculation corresponds to the state for which the orbitals have been optimized.
It is not uncommon, for instance, to find a targeted excited state lower in energy than the physical ground state,
since the former is much more correlated than the latter in the corresponding state-specific CI calculation.
We identify the state of interest by simply inspecting the coefficients of the reference determinants.
}

We can further compute the renormalized second-order Epstein-Nesbet (EN2) perturbation correction \cite{Garniron_2019} from the determinants left outside the truncated CISD space of each calculation,
giving rise to the $\Delta$CISD+EN2 model.
The EN2 perturbative correction involves a single loop over external determinants that are connected to the internal determinants via at most double excitations, thus entailing an overall $\order*{N^8}$ scaling
associated with the number of quadruply-excited determinants.
Despite this unfavorable scaling, the corresponding prefactor of the EN perturbative correction is rather small, making such calculations still affordable.
Alternatively, we could calculate one of the several types of a posteriori Davidson corrections \cite{Langhoff_1974,Pople_1977,Szalay_2012} in a state-specific fashion, leading to a $\Delta$CISD+Q approach.
We recall that computing Davidson corrections is virtually free, such that $\Delta$CISD+Q presents the same computational cost and $\order*{N^6}$ scaling as $\Delta$CISD.
\footnote{It is worth noting that the $\Delta$CSF, $\Delta$CISD, and $\Delta$CISD+Q excitation energies are invariant under rotations within the doubly-occupied and virtual blocks,
although the $\Delta$CISD+EN2 energies are not, which is well-known for the EN2 perturbative correction. \cite{Garniron_2019}}

The remaining question is how to build an appropriate reference for each state of interest.
Our general guideline is to select the smallest set of CSFs that provides a qualitatively correct description of the state of interest, as shown in Fig.~\ref{fig:determinants}.
Here we adopted the spin-restricted formalism.
The HF determinant is the obvious choice for the ground state of closed-shell singlets.
For singly-excited states of closed-shell systems, we chose either one or two CSFs, depending on each particular excited state.
For most cases, a single CSF associated with two unpaired electrons should be enough.
Some excited states, however, display a strong multireference character, like those of \ce{N2}, \ce{CO2}, and acetylene, thus requiring two CSFs.
For genuine doubly-excited states where a pair of opposite-spin electrons are promoted from the same occupied to the same virtual orbital, we selected a single determinant associated with the corresponding double excitation.
In turn, open-shell doubly-excited states were described with a single open-shell CSF (just as for most singly-excited states).
For ground and excited states of open-shell doublets, a single-determinant restricted open-shell HF reference is adopted as well.

As mentioned before, our $\Delta$CISD approach can be seen as a type of MRCI,
though with two key differences with respect to typical realizations of MRCI. \cite{Szalay_2012}
First, it relies on a minimal set of CSFs as the reference space, whereas in typical applications of MRCI the reference is built from a much larger complete active space.
This means that the CI space becomes more amenable in the former approach, enabling calculations for larger systems.
The second important difference is that the reference in $\Delta$CISD is state-specific,
which is expected to favor the overall fitness of the orbitals when compared with
state-averaged orbitals of standard MRCI (whenever excited states are involved).
$\Delta$CISD also resembles the ESMF theory \cite{Shea_2018,Shea_2020,Hardikar_2020} of Neuscamman and coworkers in their underlying motivation: a state-specific mean-field-like starting point, subject to a subsequent treatment of correlation effects.
In $\Delta$CISD, however, the starting point is much more compact and arguably closer to a mean-field description than the CIS-like ansatz of ESMF.
This makes the CI expansion up to double excitations feasible in our approach, though not in ESMF, which in turn resorts to generalized MP2 to describe correlation. \cite{Shea_2018,Clune_2020}
This $\Delta$CSF ansatz has already been suggested as a more compact alternative to the ESMF one, \cite{Shea_2020}
but again in the spirit of recovering correlation at the MP2 level, whereas we propose a state-specific CISD expansion, that could be followed by Davidson or EN2 perturbative corrections.

\section{Computational Details}
\label{sec:comp_det}

Our state-specific CI approach was implemented in {\QP}, \cite{Garniron_2019}
whose determinant-driven framework provides a very convenient platform for including arbitrary sets of determinants in the CI expansion.
In this way, we can easily select only the determinants that are connected to the reference determinants according
to a given criterion provided by the user.
On top of that, the state-specific implementation further profits from the
\textit{configuration interaction using a perturbative selection made iteratively} (CIPSI) algorithm \cite{Huron_1973,Giner_2013,Giner_2015,Garniron_2018} implemented in {\QP},
which allows for a large reduction of the CI space without loss of accuracy.
At each iteration of the CIPSI algorithm, the CI energies are obtained with the Davidson iterative algorithm, \cite{Davidson_1975}
which is ended when the EN2 perturbation correction computed in the truncated CI space lies below \SI{0.01}{\milli\hartree}. \cite{Garniron_2018}
Our state-specific CI implementation can be employed for different selection criteria for the excited determinants,
based for example on the seniority number, \cite{Bytautas_2011} the hierarchy parameter, \cite{Kossoski_2022}
or the excitation degree. Here, we considered the more traditional excitation-based CI.
After the CI calculation, we computed the renormalized EN2 perturbation correction \cite{Garniron_2019} from the determinants left outside the truncated CI space,
which is relatively cheap because of the semi-stochastic nature of our algorithm. \cite{Garniron_2017}
We also evaluate the seven variants of Davidson corrections discussed in Ref.~\onlinecite{Szalay_2012}.

To get state-specific orbitals, we first run a CIS calculation and obtained the natural transition orbitals (NTOs), \cite{Martin_2003}
which proved to be more suitable guess orbitals than the canonical HF orbitals.
The dominant hole and particle NTOs are taken as the singly-occupied orbitals,
and for pronounced multireference states, the second most important pair of NTOs was also considered (as illustrated in Fig.~\ref{fig:determinants}).
For the doubly-excited states, a non-Aufbau occupation of the canonical HF orbitals were used as guess orbitals, based on the expected character of the excitation.
The orbital optimization was performed with the Newton-Raphson method, also implemented in {\QP}. \cite{Kossoski_2021,Damour_2021}

Having our state-specific approaches presented, our main goal here is to assess their performance in describing electronic excited states.
For that, we calculated vertical excitation energies for an extensive set of 294 electronic transitions,
for systems, states, and geometries provided in the QUEST database. \cite{Veril_2021}
We considered small- \cite{Loos_2018,Loos_2021a} and medium-sized \cite{Loos_2020,Loos_2022} organic compounds,
radicals and ``exotic'' systems, \cite{Loos_2020a} and doubly-excited states. \cite{Loos_2020,Loos_2021a,Loos_2022}
The set of excited states comprises closed-shell (singlets and triplets) and open-shell (doublets) systems, ranging from one up to six non-hydrogen atoms,
and of various characters (valence and Rydberg states, singly- and doubly-excited states).
We employed the aug-cc-pVDZ basis set for systems having up to three non-hydrogen atoms, and the 6-31+G(d) basis set for the larger ones.
We compared the excitation energies obtained with our state-specific approaches against more established alternatives,
such as CIS, \cite{DelBene_1971} CIS with perturbative doubles [CIS(D)], \cite{Head-Gordon_1994,Head-Gordon_1995} CC with singles and doubles (CCSD) \cite{Purvis_1982,Scuseria_1987,Koch_1990,Stanton_1993}
and the second-order approximate CC with singles and doubles (CC2), \cite{Christiansen_1995,Hattig_2000}
the latter two understood as EOM-CC. 
The excitation energies obtained with the different methodologies were gauged against very accurate reference values,
of high-order CC or extrapolated full CI quality. \cite{Loos_2018,Loos_2021a,Loos_2020,Loos_2022,Loos_2020a,Loos_2020}
The complete set of reference methods and energies are provided in the {\SupInf}.

\section{Results and discussion}
\label{sec:res}

\subsection{Orbital optimization}

Our first important result is that the combination of the Newton-Raphson method starting with NTOs proved to be quite reliable in converging the $\Delta$CSF ansatz to excited-state solutions.
To a great extent, this is assigned to the compact reference of $\Delta$CSF, which avoids redundant solutions associated with larger MCSCF references.
In most cases, the orbitals are optimized with relatively few iterations (typically less than 10), and to the correct targeted state.
A second-order method such as Newton-Raphson is required if the targeted solution is a saddle point in the orbital rotation landscape,
which is expected to be the case for excited states. \cite{Burton_2021,Burton_2022}

At convergence, the number of negative eigenvalues of the orbital Hessian matrix, i.e., the saddle point order, can provide further insights about the topology of the solutions for a given CSF ansatz.
The full list of saddle point orders is shown in the {\SupInf}.
For the closed-shell systems, we found that the lowest-lying solution (global minimum) obtained with the open-shell CSF is always an excited state since it cannot properly describe the closed-shell character of the ground state.
In turn, higher-lying excited states tend to appear as saddle points, with increasing order as one goes up in energy, even though this behavior is not very systematic.
It was not uncommon, for example, to encounter two different excited states as local minima or sharing the same saddle point order.
For some systems, we searched for symmetry-broken solutions of excited states by rotating the orbitals along the direction associated with a negative eigenvalue of the orbital Hessian,
but this procedure leads to solutions representing a different state.
We did not explore this exhaustively though, and we cannot rule out the existence of symmetry-broken excited-state solutions.
It is also worth mentioning that the starting orbitals typically presented a much larger number of negative Hessian eigenvalues, which decreased in the course of orbital optimization.
This means that the saddle point order cannot be anticipated based on information about the unrelaxed orbitals or the expected ordering of the states.

Importantly, state-specific solutions could be found for different types of states, including singly- and doubly-excited states, for closed-shell singlets and open-shell doublets,
and for the first as well as higher-lying states of a given point group symmetry.
For this last class of states, however, our single CSF approach is not always reliable, specially for fully symmetric higher-lying states.
In some cases, a closed-shell determinant is also important (as revealed by the subsequent CISD calculation) but remains outside the open-shell CSF reference.
In these situations, employing both open- and closed-shell determinants in the reference is expected to improve the description of these higher-lying excited states,
and we plan to explore this approach in the future.
More generally, convergence issues would be expected at energies displaying a high density of excited states.

The excited-state reference could also be based on single-determinant $\Delta$SCF orbitals, rather than the $\Delta$CSF orbitals we have adopted.
However, the former method is heavily spin-contaminated, being an exact mixture of singlet and triplet, whereas the latter method targets one spin multiplicity at a time.
In this way, the excitation energies obtained with $\Delta$CSF appear above (for singlets) and below (for triplets) the single energy obtained with $\Delta$SCF,
overall improving the comparison with the reference values.
In turn, we compared $\Delta$SCF and $\Delta$CSF excited-state orbitals for $\Delta$CISD calculations, and found overall little differences in the excitation energies.
Still, we think $\Delta$CSF is preferable because it delivers truly state-specific orbitals, whereas $\Delta$SCF produces the same orbitals for the singlet and triplet states, and is thus less state-specific.

\subsection{State-specific \textit{vs} standard CI}

The state-specific $\Delta$CI approach offers a well-defined route towards full CI by increasing the excitation degree, in analogy with standard ground-state-based CI methods.
We explored both routes by calculating 16 excitation energies for small systems,
by considering up to quadruple excitations (full set of results are available in the {\SupInf}).
Even though this is a small set for obtaining significant statistics, it is enough to showcase the main trends when comparing state-specific and ground-state-based CI methods.
The mean signed error (MSE), mean absolute error (MAE), and root-mean-square error (RMSE) are shown in Table \ref{tab:DCIvsCI}.
The convergence for standard CI is quite slow,
with CISD largely overestimating the excitation energies, CISDT leading to more decent results, which are improved at the CISDTQ level.
In turn, we found that $\Delta$CI displays much more accurate results and accelerated convergence than their ground-state-based counterparts.
Already at the $\Delta$CISD level, the accuracy is far superior to that of standard CISD, being comparable to CISDT.
Going one step further ($\Delta$CISDT) does not lead to a visible improvement,
whereas the state-specific quadruple excitations of $\Delta$CISDTQ recover much of the remaining correlation energy of each state, hence the very accurate excitation energies.
These observations parallel the common knowledge that the ground state correlation energy is mostly affected by the double excitations, and that quadruples are more important than triples,
meaning that the state-specific $\Delta$CI approach manages to capture correlation effects in a reasonably balanced way for ground and excited states.
This also motivates us to investigate various flavors of Davidson correction, which attempts to capture the missing contribution from the quadruple excitations.
As will be discussed in more detail later, the popular Pople correction, \cite{Pople_1977} labeled $\Delta$CISD+PC from hereon, was found to be somewhat more accurate than the others.
The comparable MAEs of $\Delta$CISD and CISDT
can be understood from the observation that the doubly-excited determinants accessed from the excited-state reference can only be achieved via triple excitations from the ground-state reference.
The comparison between state-specific and ground-state-based CI for a given excitation degree ($\Delta$CISD against CISD and $\Delta$CISDTQ against CISDTQ)
shows that the MAEs are reduced by one order of magnitude in the former route, when compared with the latter.
No gain is observed from CISDT to $\Delta$CISDT, though.

\begin{table}[ht!]
\caption{Mean Signed Error (MSE), Mean Absolute Error (MAE), and Root-Mean Square Error (RMSE) in Units of eV, with Respect to Reference Theoretical Values,
for the Set of 16 Excitation Energies Listed in the {\SupInf}.
}
\label{tab:DCIvsCI}
\begin{ruledtabular}
\begin{tabular}{lddd}
method            &     \mc{1}{c}{MSE} & \mc{1}{c}{MAE} & \mc{1}{c}{RMSE} \\
\hline
CISD      &   +3.91 & 3.91 &  4.08 \\
CISDT     &   +0.07 & 0.17 &  0.19 \\
CISDTQ    &   +0.13 & 0.13 &  0.15 \\
\hline
$\Delta$CISD      & -0.14 & 0.18 & 0.22 \\
$\Delta$CISDT     & -0.20 & 0.20 & 0.23 \\
$\Delta$CISDTQ    & -0.02 & 0.02 & 0.03 \\
$\Delta$CISD+EN2  & -0.00 & 0.03 & 0.03 \\
$\Delta$CISD+PC   & -0.10 & 0.14 & 0.13 \\
\hline
CIS(D)            & -0.03 & 0.35 & 0.40 \\
CC2               & -0.05 & 0.32 & 0.37 \\
CCSD              & +0.01 & 0.08 & 0.09 \\
CC3               & +0.01 & 0.03 & 0.06 \\
CCSDT             & -0.01 & 0.02 & 0.02 \\
\end{tabular}
\end{ruledtabular}
\end{table}

\subsection{State-specific CI \textit{vs} other methods}

\begin{figure*}
\includegraphics[width=1.0\linewidth]{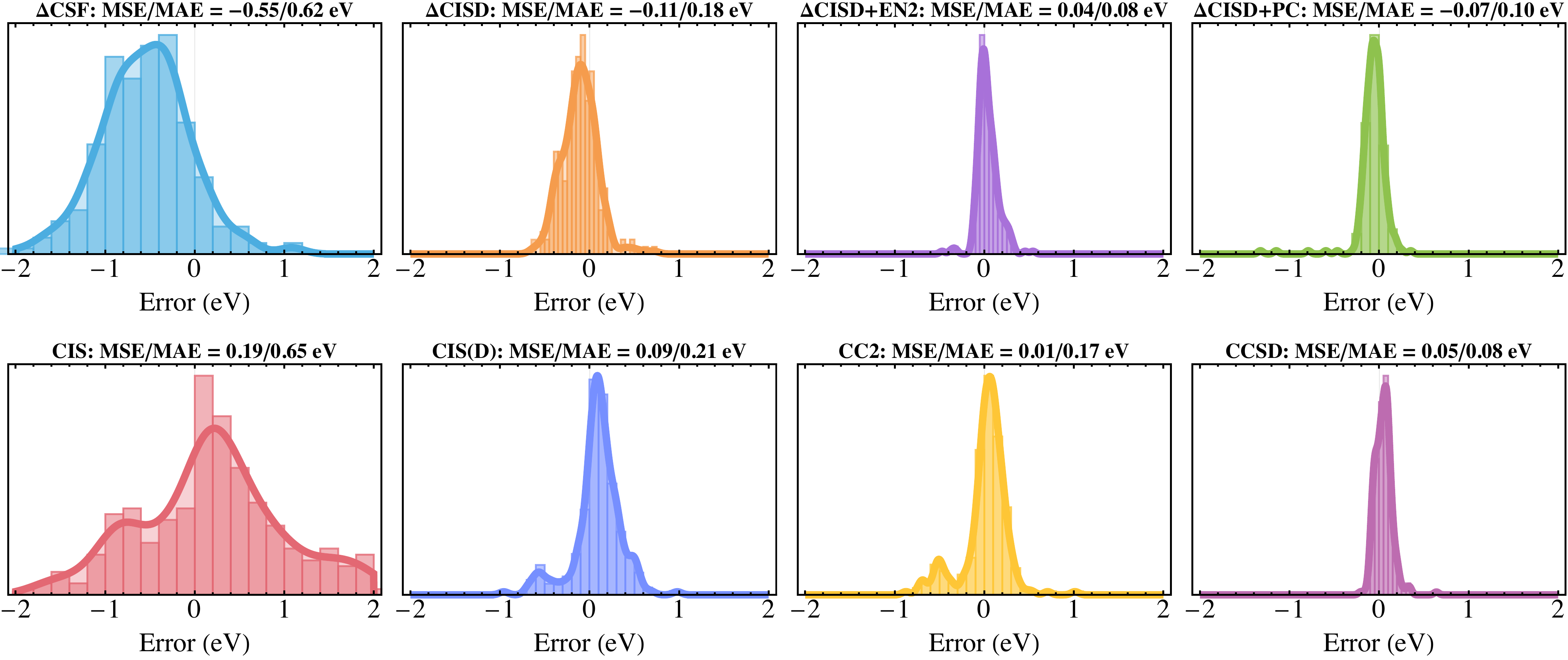}
\caption{Distribution of Errors in Excitation Energies with Respect to Reference Theoretical Values and Corresponding Mean Signed Error (MSE) and Mean Absolute Error (MAE), for Various Excited-State Methodologies.}
\label{fig:PDF}
\end{figure*}

We now start the discussion on how well our state-specific CI approaches compare with more established methods
by presenting in Fig.~\ref{fig:PDF} and in Table \ref{tab:singly} the distribution of errors and statistical measures associated with a set of 237 singly-excited states of closed-shell systems.

At the $\Delta$CSF level, the excitation energies are systematically underestimated, thus a substantially negative MSE.
\alert{
A large absolute MSE would be expected from any mean-field approach.
At least to some extent, the bias towards underestimated energies appears because the CSF reference for the excited states (typically containing two determinants) already recovers some correlation,
whereas the one-determinant HF reference of the ground state does not.
}
The MAE of the $\Delta$CSF approach (\SI{0.62}{\eV}) is comparable to that of CIS (\SI{0.65}{\eV}).
The overall similar performance of these two methods is somewhat expected since the orbital relaxation that takes place in the state-specific CSF is partially described via the single excitations of CIS.

Moving to the $\Delta$CISD level, we find that correlation effects are described in a reasonably balanced way for ground and excited states.
The MAE is significantly reduced (\SI{0.18}{\eV}) with respect to that of $\Delta$CSF,
being smaller than in CIS(D) (\SI{0.21}{\eV}) and comparable to CC2 (\SI{0.17}{\eV}).
The absolute MSE also decreases, but remains negative, whereas the other CI- or CC-based methods present positive MSEs.
This shows that there is still some bias toward a better description of excited states at the $\Delta$CISD level,
probably due to the two-determinant reference (compared to one-determinant for the ground state).
In addition, higher-lying fully symmetric states are not as well described at the $\Delta$CISD level,
reflecting the lack of a closed-shell determinant in the reference, as discussed above. We did not discard these states from the statistics though.

\begin{table}[ht!]
\caption{Mean Signed Error (MSE), Mean Absolute Error (MAE), and Root-Mean Square Error (RMSE), in Units of eV, with Respect to Reference Theoretical Values,
for the Set of 237 Excitation Energies for Singly-Excited States of Closed-Shell Systems Listed in the {\SupInf}.
}
\label{tab:singly}
\begin{ruledtabular}
\begin{tabular}{lddd}
        \mc{1}{l}{Method}            &   \mc{1}{c}{MSE} &  \mc{1}{c}{MAE} & \mc{1}{c}{RMSE} \\
\hline
$\Delta$CSF       & -0.55 & 0.62 & 0.74 \\
$\Delta$CISD      & -0.11 & 0.18 & 0.23 \\
$\Delta$CISD+EN2  & +0.04 & 0.08 & 0.12 \\
$\Delta$CISD+PC   & -0.07 & 0.10 & 0.17 \\
\hline
CIS               & +0.19 & 0.65 & 0.68 \\
CIS(D)            & +0.09 & 0.21 & 0.27 \\
CC2               & +0.01 & 0.17 & 0.25 \\
CCSD              & +0.05 & 0.08 & 0.11 \\
\end{tabular}
\end{ruledtabular}
\end{table}

\begin{figure*}
\includegraphics[width=1.0\linewidth]{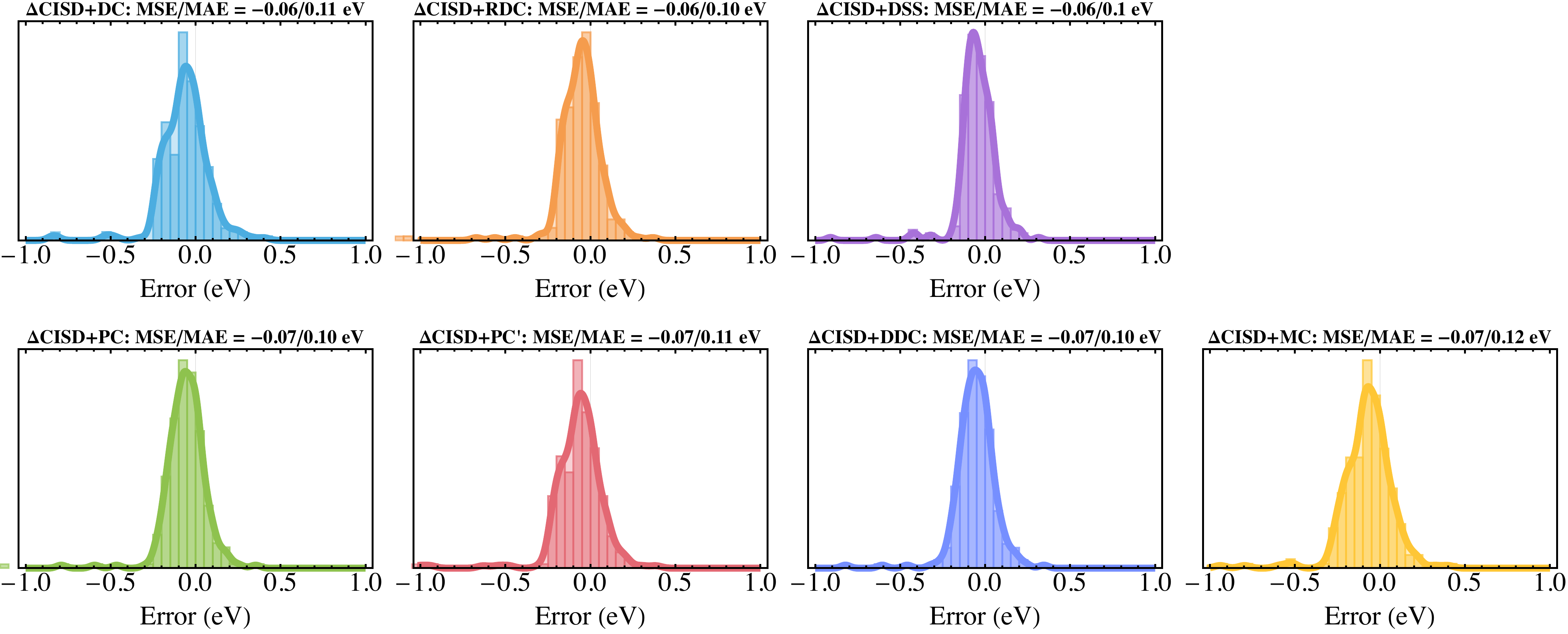}
\caption{Distribution of Errors in Excitation Energies with Respect to Reference Theoretical Values and Corresponding Mean Signed Error (MSE) and Mean Absolute Error (MAE), for Various Forms of Davidson-Corrected $\Delta$CISD+Q Models.
The Different Types of Davidson Corrections Can Be Found in Ref.~\onlinecite{Szalay_2012}.
        }
\label{fig:PDF_Q}
\end{figure*}

The perturbative correction introduced at the $\Delta$CISD+EN2 approach reduces the statistical errors even more,
showing the same MAE as CCSD (\SI{0.06}{\eV}).
At times, however, the EN2 correction leads to erroneous results due to the presence of intruder states,
which sometimes appears for the more problematic higher-lying states of a given symmetry.
We have discarded 10 out of 294 problematic cases when evaluating the statistics of the $\Delta$CISD+EN2 results.
Instead of relying on perturbation theory to correct the CISD total energies, we can resort to one of Davidson corrections. \cite{Szalay_2012}
Even though this correction is not as accurate as the EN2 perturbative energy,
more often than not it improves upon $\Delta$CISD, and with virtually no additional computational cost.
For the $\Delta$CISD+Q statistics, we discarded 12 out of 294 data points where $\norm{\boldsymbol{c}} < 0.9$, where $\boldsymbol{c}$ gathers the coefficients of the reference determinants in the CI expansion.
We found that all seven $\Delta$CISD+Q variants provide MAEs in the \SIrange{0.10}{0.12}{\eV} range,
with the individual distribution of errors and statistical measures presented in Fig.~\ref{fig:PDF_Q}.
As alluded to before, the Pople corrected flavor, $\Delta$CISD+PC, is arguably the most well-behaved one, with fewer outlier excitation energies and the lowest MAEs, of \SI{0.10}{\eV}.

We also surveyed the performance of our state-specific methods for 10 genuine doubly-excited states \cite{Loos_2019} and 47 excited states of open-shell doublets (doublet-doublet transitions), \cite{Loos_2020a}
both sets were extracted from the QUEST database. \cite{Veril_2021}
The statistical measures are shown in Table \ref{tab:all_sets}, together with those of singly-excited states of closed-shell systems.
The important finding in this comparison is that the state-specific methods provide reasonably similar MAEs for the three types of excited states.
For instance, $\Delta$CISD has MAEs of \SI{0.18}{\eV} for singly-excited states of closed-shell singlets, \SI{0.17}{\eV} for doublet-doublet transition and \SI{0.16}{\eV} for doubly-excited states.
This contrasts with the case of more familiar methods, which cannot describe doubly-excited states unless higher-order excitations are included. \cite{Loos_2019}
We notice that the MSE of $\Delta$CSF is more negative for singly-excited states of closed-shell molecules (\SI{-0.55}{\eV}) than for doubly-excited states (\SI{-0.20}{\eV}),
being closer to zero for doublet-doublet transitions (\SI{+0.07}{\eV}),
which reflects the one-determinant reference adopted for both the excited and ground states in the latter cases.
This difference does not translate into comparatively smaller errors in the correlated results though.

\begin{table*}[ht!]
\caption{Mean Signed Error (MSE), Mean Absolute Error (MAE), and Root-Mean Square Error (RMSE), in Units of eV, with Respect to Reference Theoretical Values,
for the Excitation Energies of 237 Singly-Excited States (Set A) and 10 Doubly-Excited States (Set B) from Closed-Shell Singlets, and of 47 Singly-Excited States from Open-Shell Doublets (Set C) Listed in the {\SupInf}.
}
\label{tab:all_sets}
\begin{ruledtabular}
\begin{tabular}{lddddddddddddd}
        &   & \multicolumn{3}{c}{$\Delta$CSF} & \multicolumn{3}{c}{$\Delta$CISD} & \multicolumn{3}{c}{$\Delta$CISD+EN2} & \multicolumn{3}{c}{$\Delta$CISD+PC} \\
\cline{3-5} \cline{6-8} \cline{9-11} \cline{12-14}
         &  \mc{1}{c}{\# States}      &     \mc{1}{c}{MSE} &  \mc{1}{c}{MAE} & \mc{1}{c}{RMSE}   &      \mc{1}{c}{MSE} &  \mc{1}{c}{MAE} &\mc{1}{c}{RMSE}   &      \mc{1}{c}{MSE} &  \mc{1}{c}{MAE} & \mc{1}{c}{RMSE}    &     \mc{1}{c}{MSE} &  \mc{1}{c}{MAE} &\mc{1}{c}{RMSE} \\
\hline
Set A         &   237   &   -0.55 & 0.62 & 0.74    &   -0.11 & 0.18 & 0.23    &   +0.04 & 0.08 & 0.12    &   -0.07 & 0.10 & 0.17 \\
Set B         &    10   &   -0.20 & 0.46 & 0.64    &   +0.03 & 0.16 & 0.22    &   -0.05 & 0.13 & 0.16    &   -0.10 & 0.13 & 0.16 \\
Set C         &    47   &   +0.07 & 0.49 & 0.64    &   +0.07 & 0.17 & 0.25    &   -0.02 & 0.07 & 0.11    &   -0.03 & 0.09 & 0.16 \\
\hline
All sets      &   294   &   -0.44 & 0.61 & 0.72    &   -0.07 & 0.17 & 0.23    &   +0.02 & 0.08 & 0.12    &   -0.06 & 0.10 & 0.17 \\
\end{tabular}
\end{ruledtabular}
\end{table*}

For the doubly-excited states, we further compare in Table \ref{tab:doubly} the performance of state-specific CI against higher-order CC methods.
The accuracy of the $\Delta$CSF mean-field model is superior to CC3 and approaches CCSDT, which highlights the importance of orbital relaxation for doubly-excited states.
$\Delta$CISD is significantly more accurate than CCSDT, whereas the perturbative and the Davidson corrections bring a small improvement.
It is worth mentioning recent developments and promising results with
state-specific CC \cite{Lee_2019,Levi_2020,Kossoski_2021,Marie_2021} and DFT \cite{Hait_2020,Hait_2021} for doubly-excited states.
However, these approaches are still restricted to states dominated by a single closed-shell determinant,
whereas the $\Delta$CISD approach can handle both closed- and open-shell doubly-excited states.
Out of the 10 doubly-excited states we investigated, only 5 (beryllium, ethylene, formaldehyde, nitroxyl, and nitrosomethane) can be qualitatively described with a single closed-shell determinant,
whereas at least two determinants are needed for the remaining 5 states:
two closed-shell determinants for glyoxal and carbon dimer (\ce{C2}), and four closed-shell determinants for carbon trimer (\ce{C3}).

\begin{table}[ht!]
\caption{Mean Signed Error (MSE) and Mean Absolute Error (MAE), in Units of eV, with Respect to Reference Theoretical Values,
for the Set of 10 Doubly-Excited States Listed in the {\SupInf}.
}
\label{tab:doubly}
\begin{ruledtabular}
\begin{tabular}{ldd}
Method            &     \mc{1}{c}{MSE} & \mc{1}{c}{MAE} \\
\hline
$\Delta$CSF       &   -0.20 & 0.46 \\
$\Delta$CISD      &   +0.03 & 0.16 \\
$\Delta$CISD+EN2  &   -0.05 & 0.13 \\
$\Delta$CISD+PC   &   +0.02 & 0.13 \\
\hline
CC3               &   +0.85 & 0.85 \\
CCSDT             &   +0.38 & 0.38 \\
CCSDTQ            &   +0.03 & 0.03 \\
\end{tabular}
\end{ruledtabular}
\end{table}

\subsection{Types of excitation}

The performance of our state-specific methods can also be assessed for specific types of excited states, say for $\pi\pi^*$ transitions or for systems of a given size.
This is shown in Table \ref{tab:mae_singly}, which compares the MAEs across different categories,
whereas the corresponding MSEs and RMSEs can be found in Tables S1 and S2 of the {\SupInf}.
Many trends can be identified, but here we highlight the most notorious and interesting ones.

Starting with spin multiplicity,
we found that the $\Delta$CISD results are comparable for singlets and triplets,
whereas the perturbative correction has a more pronounced effect for the triplets,
bringing the MAE down to \SI{0.06}{\eV}, the same as obtained with CCSD.
To some extent, the worse performance of the EN2 correction for the singlets stems from intruder states
(most noticeably the ground state when it shares the same point group symmetry as the targeted excited state).
We also found that the Davidson corrections bring a somewhat larger improvement for triplets than for singlets,
with some flavors having MAEs of \SI{0.07}{\eV} for the triplets, being essentially as accurate as CCSD (MAE of \SI{0.06}{\eV}).
For the doublet-doublet transitions, the EN2 and +Q corrections are as helpful as they are for the triplets (see Table \ref{tab:all_sets} and \ref{tab:mae_singly}).

\begin{table*}[ht!]
\caption{Mean Absolute Error, in Units of eV, with Respect to Reference Theoretical Values,
for Different Types of Singly-Excited States of Closed-Shell Systems Listed in the {\SupInf}.
}
\label{tab:mae_singly}
\begin{ruledtabular}
\begin{tabular}{lddddddddd}
& \mc{1}{c}{\# States} & \mc{1}{c}{$\Delta$CSF} & \mc{1}{c}{$\Delta$CISD} & \mc{1}{c}{$\Delta$CISD+EN2} & \mc{1}{c}{$\Delta$CISD+PC} & \mc{1}{c}{CIS} & \mc{1}{c}{CIS(D)} & \mc{1}{c}{CC2} & \mc{1}{c}{CCSD} \\
\hline
All states      & 237 & 0.62 & 0.18 & 0.08 & 0.10 & 0.65 & 0.21 & 0.17 & 0.08 \\
\hline
Singlet         & 127 & 0.56 & 0.17 & 0.10 & 0.12 & 0.68 & 0.22 & 0.19 & 0.10 \\
Triplet         & 110 & 0.69 & 0.18 & 0.06 & 0.09 & 0.61 & 0.19 & 0.15 & 0.06 \\
\hline
Valence         & 155 & 0.65 & 0.21 & 0.08 & 0.10 & 0.61 & 0.19 & 0.14 & 0.08 \\
Rydberg         &  82 & 0.56 & 0.12 & 0.10 & 0.11 & 0.72 & 0.24 & 0.22 & 0.08 \\
\hline
$n\pi^*$        &  56 & 0.60 & 0.16 & 0.06 & 0.07 & 0.54 & 0.10 & 0.09 & 0.06 \\
$\pi\pi^*$      &  80 & 0.75 & 0.26 & 0.10 & 0.12 & 0.74 & 0.27 & 0.20 & 0.10 \\
$\sigma\pi^*$   &  18 & 0.39 & 0.13 & 0.03 & 0.06 & 0.26 & 0.11 & 0.07 & 0.04 \\
$n$   Rydberg   &  40 & 0.61 & 0.12 & 0.12 & 0.14 & 1.17 & 0.37 & 0.39 & 0.07 \\
$\pi$ Rydberg   &  38 & 0.46 & 0.10 & 0.08 & 0.09 & 0.29 & 0.11 & 0.06 & 0.10 \\
\hline
1-2 non-H atoms &  69 & 0.83 & 0.18 & 0.06 & 0.12 & 0.71 & 0.24 & 0.23 & 0.06 \\
3-4 non-H atoms & 122 & 0.57 & 0.18 & 0.09 & 0.10 & 0.70 & 0.20 & 0.16 & 0.08 \\
5-6 non-H atoms &  46 & 0.41 & 0.15 & 0.12 & 0.07 & 0.43 & 0.18 & 0.12 & 0.10 \\
\end{tabular}
\end{ruledtabular}
\end{table*}

Regarding the character of the excitations, we found that $\Delta$CISD is considerably better for Rydberg (MAE of \SI{0.12}{\eV}) than for valence (MAE of \SI{0.21}{\eV}) excited states.
In turn, the EN2 correction has a larger impact on valence excitations, making little difference for the Rydberg states, such that the $\Delta$CISD+EN2 results become comparable for both types of excitation,
with MAEs of \SIrange{0.08}{0.10}{\eV}.
Additional trends can be observed when dividing the valence excitations into $n\pi^*$, $\pi\pi^*$, or $\sigma\pi^*$,
and the Rydberg excitations as taking place from $n$ or $\pi$ orbitals.
Our state-specific methods are typically more accurate for $n\pi^*$ excitations than for $\pi\pi^*$ excitations.
$\Delta$CISD+EN2, for example, is as accurate as CCSD for $n\pi^*$ transitions, with corresponding MAEs of \SI{0.06}{\eV}.
We also found that the less common $\sigma\pi^*$ excitations are much better described across all methods than the more typical $n\pi^*$ and $\pi\pi^*$ transitions.
For this type of state, $\Delta$CISD+EN2 is the best performing method, with MAEs as small as \SI{0.03}{\eV}.
When separating the Rydberg states by the character of the hole orbital, $n$ or $\pi$, additional interesting features can be seen.
Except for CCSD, all the other methods considered here provide more accurate results for the Rydberg excitations involving the $\pi$ orbitals.
Not only that, but the MAEs are quite small and comparable across all methods (except for $\Delta$CSF and CIS), ranging from \SIrange{0.06}{0.11}{\eV}.
Surprisingly, CIS is much more accurate for $\pi$ Rydberg (MAE of \SI{0.29}{\eV}) than for $n$ Rydberg (MAE of \SI{1.17}{\eV}) excitations.

The third and most important line of comparison concerns the system size.
Under this criterion, we divided the excited states into three groups, small, medium, and large, depending on the number of non-hydrogen atoms (see Table \ref{tab:mae_singly}).
We found that $\Delta$CSF becomes more accurate as the system size increases,
which we assign to a diminishing effect of the one- \textit{vs} two-determinant imbalance discussed above.
As the system size increases, the correlation energy recovered by the two determinants of the excited states (at the $\Delta$CSF level) is expected to become smaller in comparison to the total correlation energy (associated with the full Hilbert space), thus alleviating this imbalance.
In contrast, CISD should provide less accurate total energies for larger systems, due to its well-known lack of size-consistency.
This issue would be expected to reflect on excitation energies to some degree, which are not absolute but relative energies.
However, a more balanced reference provided by $\Delta$CSF might compensate for the lack of size-consistency when larger systems are targeted.
Indeed, $\Delta$CISD presents comparable MAEs across the three sets of system size (\SIrange{0.15}{0.18}{\eV}).
In contrast, $\Delta$CISD+Q and $\Delta$CISD+EN2 seem to go opposite ways: the former becomes more accurate and the latter less as a function of system size.
Similarly, CC2 becomes more accurate and CCSD loses accuracy as the system size increases, \cite{Loos_2020,Loos_2021}
to the point where the theoretically more approximate CC2 becomes the favored methodological choice.
It remains to be seen how the absence of size-consistency in $\Delta$CISD impairs the results for even larger systems than those considered here,
and the extent to which Davidson or perturbative corrections reduce this problem.
For molecules containing five or six non-hydrogen atoms, $\Delta$CISD+EN2 becomes practically as accurate as CCSD, with MAEs in the \SIrange{0.10}{0.12}{\eV} range.
The $\Delta$CISD+Q models turn out to be the most accurate choice for systems of this size,
with MAEs ranging from \SIrange{0.07}{0.09}{\eV} (see Table S3 of the {\SupInf}),
$\Delta$CISD+PC displaying a MAE of only \SI{0.07}{\eV}.
In particular, it is more accurate than CCSD while sharing the same $\order*{N^6}$ computational scaling,
and more accurate than CC2, despite remaining less black-box and more expensive than the $\order*{N^5}$ scaling of the latter.
Overall, the present statistics place our state-specific approaches as encouraging alternatives for describing larger systems,
despite the remaining issues regarding higher-lying excited states.
The MAEs of the seven variants of Davidson corrections, separated by type of excitation, are presented in Table S3 of the {\SupInf}.
We recall that different basis have been used, the aug-cc-pVDZ basis set for systems with up to three non-hydrogen atoms, and the 6-31+G(d) basis set for the larger ones,
which could have some impact on the trends as a function of system size, for a given method.
Despite the different basis sets, the comparison between different methods, for a given system size, remains valid.

\subsection{Specific applications}

Butadiene, glyoxal, \ce{C2} and \ce{C3} are particularly interesting and challenging systems that deserve a dedicated discussion.
The excitation energies are gathered in Tables \ref{tab:butadiene}, \ref{tab:glyoxal}, \ref{tab:c2}, and \ref{tab:c3}, respectively.

\begin{table}[ht!]
\caption{Excitation Energies of Butadiene, in Units of eV, According to Different Methodologies.
The reference method is CCSDTQ \cite{Loos_2022}.
Only the lowest-lying optically bright ($1 {}^1 B_u$) and dark ($2 {}^1 A_g$) states and their energy gap are compared here.
Seven more states have been computed, which can also be found in the {\SupInf}.
}
\label{tab:butadiene}
\begin{ruledtabular}
\begin{tabular}{lddddd}
        \mc{1}{l}{Method} & \mc{1}{c}{$1 {}^1 B_u$} & \mc{1}{c}{$2 {}^1 A_g$} & \mc{1}{c}{MSE} & \mc{1}{c}{MAE} & \mc{1}{c}{$1 {}^1 B_u$/$2 {}^1 A_g$ gap} \\
\hline
$\Delta$CSF       & 6.53 & 7.18        & +0.37 & 0.37 &  0.65 \\
$\Delta$CISD      & 6.54 & 6.93        & +0.25 & 0.25 &  0.39 \\
$\Delta$CISD+EN2  & 6.51 & 5.59\fnm[1] &   -   &   -  &   -   \\
$\Delta$CISD+PC   & 6.48 & 1.61\fnm[1] &   -   &   -  &   -   \\
\hline
CIS               & 6.30 & 7.56 & +0.44 & 0.44 & 1.25 \\
CIS(D)            & 6.15 & 7.53 & +0.49 & 0.49 & 1.12 \\
CC2               & 6.32 & 7.26 & +0.31 & 0.40 & 0.94 \\
CCSD              & 6.55 & 7.20 & +0.39 & 0.39 & 0.65 \\
\hline
Reference method  & 6.41 & 6.56 &   -   & - & 0.15 \\
\end{tabular}
\end{ruledtabular}
\fnt[1]{Intruder state problem for this state.}
\end{table}

The dark $2 {}^1 A_{g}$ excited state of butadiene is a notoriously famous example, having received much attention (see Ref.~\onlinecite{Loos_2019} and references therein).
Prior studies had assigned it as a doubly-excited state, due to important contributions from doubly-excited determinants. \cite{Serrano-Andres_1993,Starcke_2006}
More recently, though, it has been re-assigned as a singly-excited state, \cite{Barca_2018a}
meaning that the doubly-excited determinants actually represent strong orbital relaxation effects
(single excitations from the dominant singly-excited determinant).
Here, our state-specific results (shown in Table \ref{tab:butadiene}) support this interpretation, since one CSF associated with a single excitation provided reasonable excitation energies,
whereas attempts to employ a doubly-excited reference produced much higher-lying solutions.
Already at the $\Delta$CSF level, we obtained an excitation energy (\SI{7.18}{\eV}) comparable to the much more expensive CCSD (\SI{7.20}{\eV}), though still overestimating the CCSDTQ reference value of \SI{6.56}{\eV}. \cite{Loos_2020}
This result demonstrates the ability of $\Delta$CSF to capture orbital relaxation effects, at only a mean-field cost, which in contrast needs at least double excitations in EOM-CC.
Inclusion of correlation at the $\Delta$CISD level brings the excitation energy down to \SI{6.93}{\eV}.
An important question in butadiene concerns the energy gap between the $2 {}^1 A_g$ dark state and the lower-lying $1 {}^1 B_u$ bright state, whose correct ordering has only recently been settled. \cite{Watson_2012}
Having the CCSDTQ reference value of \SI{0.15}{\eV} for the energy gap, \cite{Loos_2020}
we observe that EOM-CC methods considerably overestimate it (\SI{0.94}{\eV} in CC2 and \SI{0.65}{\eV} in CCSD),
whereas the state-specific methods deliver improved results (\SI{0.65}{\eV} in $\Delta$CSF and \SI{0.39}{\eV} in $\Delta$CISD).

\begin{table*}[htb!]
\caption{Excitation Energies of Glyoxal, in Units of eV, According to Different Methodologies.
The number in parenthesis represents the number of CSFs considered in the reference.
The reference method is CCSDTQ for the singlets \cite{Loos_2022} and CCSDT for the triplets. \cite{Loos_2020}
}
\label{tab:glyoxal}
\begin{ruledtabular}
\begin{tabular}{ldddddddd}
        \mc{1}{l}{Method}& \mc{1}{c}{$1 {}^1 A_u$} & \mc{1}{c}{$1 {}^1 B_g$} & \mc{1}{c}{$1 {}^3 A_u$} & \mc{1}{c}{$1 {}^3 B_g$} & \mc{1}{c}{$1 {}^3 B_u$} & \mc{1}{c}{$1 {}^3 A_g$} & \mc{1}{c}{MSE} & \mc{1}{c}{MAE} \\
\hline
$\Delta$CSF(1)       & 4.11 & 5.88 & 3.53 & 5.50 & 4.69 & 7.44 & +0.97 & 1.14 \\
$\Delta$CSF(2)       & 3.34 & 4.56 & 2.70 & 3.91 & 3.86 & 5.06 & -0.31 & 0.58 \\
$\Delta$CISD(1)      & 3.49 & 5.27 & 3.00 & 4.85 & 5.37 & 7.23 & +0.65 & 0.65 \\
$\Delta$CISD(2)      & 3.12 & 4.51 & 2.59 & 3.97 & 4.75 & 5.90 & -0.08 & 0.22 \\
$\Delta$CISD(1)+EN2  & 3.16 & 4.86 & 2.80 & 4.46 & 5.57 & 7.00 & +0.42 & 0.42 \\
$\Delta$CISD(2)+EN2  & 3.17 & 4.57 & 2.61 & 4.08 & 5.10 & 6.27 & +0.08 & 0.15 \\
$\Delta$CISD(1)+PC   & 3.10 & 4.72 & 2.70 & 4.32 & 5.35 & 6.74 & +0.27 & 0.27 \\
$\Delta$CISD(2)+PC   & 2.92 & 4.33 & 2.50 & 3.91 & 5.03 & 6.19 & -0.07 & 0.08 \\
\hline
Reference method     & 2.94 & 4.31 & 2.55 & 3.95 & 5.20 & 6.35 &  -    & -    \\
\end{tabular}
\end{ruledtabular}
\end{table*}

Another challenging system is glyoxal, which presents excited states of genuine multireference character. \cite{Hollauer_1991}
While the first pair of NTOs has a dominant weight, the second pair is non-negligible.
In this sense, most of the first excited states of glyoxal lie in between the cases of most singly-excited states (that can be qualitatively described with one CSF) and those that need two CSFs.
Being an intermediate case, here we performed $\Delta$CISD calculations with references containing one or two CSFs, for the first two singlet states and first four triplet states
(results presented in Table \ref{tab:glyoxal}).
With one CSF only, $\Delta$CSF typically overestimates the reference excitation energies, with the corresponding $\Delta$CISD improving the overall comparison.
For this set of six excited states, the associated MAEs are \SI{1.14}{\eV} for $\Delta$CSF and \SI{0.65}{\eV} for $\Delta$CISD when using a single CSF as reference.
Despite the improvement brought at the CISD level, this is still limited by the lack of an actual multiconfigurational reference for these states.
When two CSFs are employed as the reference for the excited states, the MAEs are reduced to \SI{0.58}{\eV} ($\Delta$CSF) and \SI{0.22}{\eV} ($\Delta$CISD),
which can be further decreased to \SI{0.08}{\eV} with $\Delta$CISD+PC.
We thus recommend augmenting the excited-state reference whenever it displays at least some multireference character,
and the weight of the first pairs of NTOs could serve as an easy proxy for this.

\begin{table}[ht!]
\caption{Excitation Energies of \ce{C2}, in Units of eV, According to Different Methodologies.
The coupled-cluster and reference (extrapolated full configuration interaction) results are from Ref.~\onlinecite{Loos_2021a}.
}
\label{tab:c2}
\begin{ruledtabular}
\begin{tabular}{ldddd}
        \mc{1}{l}{Method} & \mc{1}{c}{$1 {}^1 \Delta_g$} & \mc{1}{c}{$2 {}^1 \Sigma_g^+$} & \mc{1}{c}{MSE} & \mc{1}{c}{MAE} \\
\hline
$\Delta$CSF       &  0.83  & 1.37 & -1.26 & 1.26   \\
$\Delta$CISD      &  1.80  & 2.33 & -0.29 & 0.29  \\
$\Delta$CISD+EN2  &  2.16  & 2.35 & -0.10 & 0.10  \\
$\Delta$CISD+PC   &  2.09  & 2.18 & -0.24 & 0.24  \\
\hline
CC3               &  3.11  & 3.28 & +0.84 & 0.84  \\
CCSDT             &  2.63  & 2.87 & +0.40 & 0.40  \\
CC4               &  2.34  & 2.60 & +0.11 & 0.11  \\
CCSDTQ            &  2.24  & 2.52 & +0.02 & 0.02  \\
\hline
Reference method  &  2.21  & 2.50 & -  & -   \\
\end{tabular}
\end{ruledtabular}
\end{table}

\begin{table}[ht!]
\caption{Excitation Energies of \ce{C3}, in Units of eV, According to Different Methodologies.
The coupled-cluster and reference (extrapolated full configuration interactions) results are from Ref.~\onlinecite{Loos_2019}.
}
\label{tab:c3}
\begin{ruledtabular}
\begin{tabular}{ldddd}
        \mc{1}{l}{Method}& \mc{1}{c}{$1 {}^1 \Delta_g$} & \mc{1}{c}{$2 {}^1 \Sigma_g^+$} & \mc{1}{c}{MSE} & \mc{1}{c}{MAE} \\
\hline
$\Delta$CSF       &  5.10  & 5.88        & -0.06 & 0.06 \\
$\Delta$CISD      &  5.17  & 5.93        & +0.01 & 0.04 \\
$\Delta$CISD+EN2  &  5.29  & 5.29\fnm[1] &   -   &  -   \\
$\Delta$CISD+PC   &  5.12  & 4.19\fnm[1] &   -   &  -   \\
\hline
CC3               &  6.65  & 7.20 & +1.38 & 1.38 \\
CCSDT             &  5.82  & 6.49 & +0.61 & 0.61 \\
CCSDTQ            &  5.31  & 6.00 & +0.11 & 0.11 \\
\hline
Reference method  &  5.21  & 5.88 & - & -   \\
\end{tabular}
\end{ruledtabular}
\fnt[1]{Intruder state problem for this state.}
\end{table}

Finally, we comment on the lowest-lying $1 ^1 \Delta_g$ and higher-lying $2 ^1 \Sigma_g^+$ doubly-excited states of \ce{C2} and \ce{C3},
which would require at least CCSDTQ quality calculations to become accurate to within \SI{0.1}{\eV}. \cite{Loos_2019}
\ce{C2} displays a strong multireference ground state, and thus we employed two CSFs as the reference,
the closed-shell HF and the determinant associated with the $(\sigma_{2s}^*)^2 \to (\sigma_{2p_z})^2$ transition.
For its doubly-excited states, we employed the two CSFs needed to describe both doubly-excited states,
generated from the HF determinant through the $(\pi_{2p_x})^2 \to (\sigma_{2p_z})^2$ and $(\pi_{2p_y})^2 \to (\sigma_{2p_z})^2$ excitations, $\pi_{2p_x}$ and $\pi_{2p_y}$ being degenerate orbitals.
In \ce{C3}, the multireference character of the ground state is weaker, and thus we adopted a single HF determinant as a reference.
In turn, four CSFs are needed for its doubly-excited states,
built from the HF determinant by performing $(\sigma_g)^2 \to (\pi_{2p_x}^*)^2$, $(\sigma_g)^2 \to (\pi_{2p_y}^*)^2$, $(\sigma_u)^2 \to (\pi_{2p_x}^*)^2$, and $(\sigma_u)^2 \to (\pi_{2p_y}^*)^2$ transitions,
where $\pi_{2p_x}^*$ and $\pi_{2p_y}^*$ are degenerate orbitals.
We therefore re-assign the doubly-excited states of \ce{C3} as $(\sigma)^2 \to (\pi^*)^2$, which had been first assigned as $(\pi)^2 \to (\sigma^*)^2$. \cite{Loos_2019}
Notice that, for both systems, what differentiates $1 ^1 \Delta_g$ and $2 ^1 \Sigma_g^+$ is essentially the phase between the two CSFs differing by the occupation of the degenerate orbitals ($\pi$ in \ce{C2}, $\pi^*$ in \ce{C3}).
Thus, the higher-lying state orbitals were obtained by optimizing for the second CI root associated with the reference (two CSFs in \ce{C2}, four in \ce{C3}).
The computed excitation energies of \ce{C2} and \ce{C3} are shown in Table \ref{tab:c2} and \ref{tab:c3}, respectively.
We found that $\Delta$CISD is more accurate than CCSDT for \ce{C2},
and even more accurate than CCSDTQ for \ce{C3}.

\section{Conclusions}
\label{sec:ccl}

To summarize, here we have presented and benchmarked a particular state-specific realization of MCSCF and MRCI
as a route to perform excited-state calculations.
The orbitals are optimized for a targeted state with a minimal set of CSFs, serving as the reference wave function for the CI calculations,
which can be further corrected with Epstein-Nesbet perturbation theory or with a posteriori Davidson corrections.
We have surveyed these methods against more established alternatives by computing excitation energies for a vast class of molecules and types of excitations from the QUEST database.
State-specific CI was found to be substantially more accurate than the standard CI methods based on a ground-state reference.
Importantly, it delivers reliable results across different types of excited states, most notably when comparing singly- and doubly-excited states, and can easily handle ground and excited states of multireference nature.
The overall accuracy of $\Delta$CISD rivals that of CC2 (MAEs of \SIrange{0.17}{0.18}{\eV}),
whereas $\Delta$CISD+EN2 is comparable to CCSD (MAEs of \SI{0.08}{\eV}),
with $\Delta$CISD+Q lying in-between (MAE of \SIrange{0.10}{0.12}{\eV}).
For larger systems, $\Delta$CISD+Q leads to more accurate results (MAE of \SIrange{0.07}{0.09}{\eV})
than CC2 and CCSD (MAEs of \SIrange{0.10}{0.12}{\eV}).

There are many exciting possibilities to be pursued from this work.
One is to develop analogous state-specific coupled-cluster methods.
In light of the huge improvement we have observed when going from ground-state-based to state-specific CI,
we expect a similar gain when comparing EOM-CC to state-specific CC methods where tailored CSFs are employed as the reference wave function. \cite{Piecuch_1992,Piecuch_1994,Lutz_2018}
One could also develop state-specific implementations of seniority-based \cite{Bytautas_2011} and hierarchy-based \cite{Kossoski_2022} CI for excited states.
It would be important to assess the performance of our state-specific approaches to
charge-transfer states
and even larger systems, which would require switching from a determinant-driven to an integral-driven implementation.
\alert{In addition, it remains to be seen how the methods presented here behave out of the equilibrium geometry, particularly in strong correlation regimes.}
Although evaluating nonorthogonal matrix elements is more challenging than their orthogonal analogs, the calculation of static properties such as dipole moments and oscillator strengths is possible thanks to the recent generalized extension of the nonorthogonal Wick's theorem proposed by Burton. \cite{Burton_2021a,Burton_2022a}
Yet another exciting possibility is to move from a state-specific to a state-averaged reference,
while contemplating only a small set of important determinants for describing a given set of states.
\alert{
We recall that the very compact reference wave function employed here is what currently limits the $\Delta$CI method to relatively low-lying excited states.
For instance, missing important determinants in the reference space give rise to intruder states encountered in some of the present $\Delta$CISD+EN2 calculations.
In particular, including the Aufbau closed-shell determinant in the reference should improve the case of fully symmetric excited states.
More generally, when two states of the same symmetry are strongly coupled, a larger reference should be considered as well.
These issues are expected to become more prominent at higher energies, due to an increasing number of excited states.
Developments toward more suitable reference wave functions could enable the $\Delta$CI method to target higher-lying excited states.
}


\begin{acknowledgements}
This work was performed using HPC resources from CALMIP (Toulouse) under allocation 2021-18005.
This project has received funding from the European Research Council (ERC) under the European Union's Horizon 2020 research and innovation programme (Grant agreement No.~863481).
\end{acknowledgements}

\section*{Supporting information available}
\label{sec:SI}

Additional statistical measures for different sets of excited states and for all flavors of $\Delta$CISD+Q models.
For the full set of 294 excited states,
total energies and excitation energies obtained at the $\Delta$CSF, $\Delta$CISD, $\Delta$CISD+EN2, and seven variants of $\Delta$CISD+Q models,
excitation energies computed with CIS, CIS(D), CC2, and CCSD,
number of determinants in the reference, saddle point order associated with the $\Delta$CSF solutions,
the reference excitation energies and corresponding method,
and additional statistical measures.
For a subset of 16 excited states, total energies and excitation energies obtained at the $\Delta$CISDT, $\Delta$CISDTQ and ground-state-based CISDT and CISDTQ levels of theory.
For the subset of 10 doubly-excited states, excitation energies obtained at the CC3, CCSDT, CC4, and CCSDTQ levels of theory.



\bibliography{manuscript}

\begin{figure*}
\begin{center}
        \boxed{
		\includegraphics[keepaspectratio,width=3.25in]{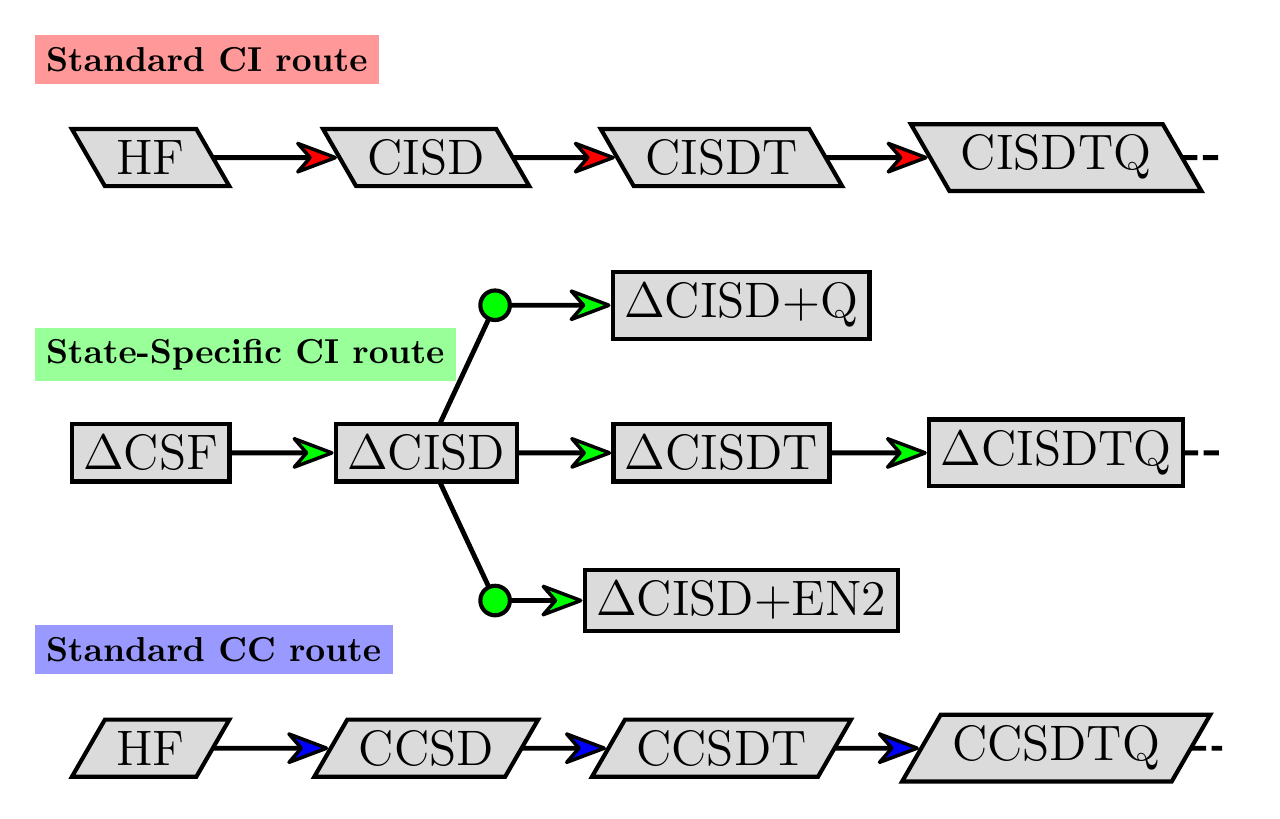}
	}
	\\
	For Table of Contents Only
\end{center}
\end{figure*}

\end{document}